\documentclass[prb,twocolumn]{revtex4}
\usepackage{graphicx}
\begin{document}

\title{Diffusive Thermal Dynamics for the Ising Ferromagnet}
\author{P. Buonsante$^{1,2,3,}$\footnote{{\tt buonsante@polito.it}},
R. Burioni$^{1,2,}$\footnote{{\tt burioni@fis.unipr.it}},
D. Cassi$^{1,2,}$\footnote{{\tt cassi@fis.unipr.it}} and
A. Vezzani$^{1,2,}$\footnote{{\tt vezzani@fis.unipr.it}}}
\address{$^1$ Istituto Nazionale Fisica della Materia.\\
$^2$ Dipartimento di Fisica, Universit\`a degli Studi di Parma,
Parco Area delle Scienze 7/a - 43100 Parma, Italy.\\
$^3$ Dipartimento di Fisica, Politecnico di Torino, Corso Duca
degli Abruzzi 24, 10129 - Torino, Italy.}
\date{\today}
\begin{abstract}
We introduce a thermal dynamics for the Ising ferromagnet where
the energy variations occurring within the system exhibit a
diffusive character typical of thermalizing agents such as e.g.
localized excitations. Time evolution is provided by a walker
hopping across the sites of the underlying lattice according to
local probabilities depending on the usual Boltzmann weight at a
given temperature. Despite the canonical hopping probabilities the
walker drives the system to a stationary state which is not
reducible to the canonical equilibrium state in a trivial way. The
system still exhibits a magnetic phase transition occurring at a
finite value of the temperature larger than the canonical one. The
dependence of the model on the density of walkers realizing the
dynamics is also discussed. Interestingly the differences between
the stationary state and the  Boltzmann equilibrium state decrease
with increasing number of walkers.
\end{abstract}
\maketitle

\hspace{.5cm}

\section{Introduction and Motivations}
Consider an Ising ferromagnet consisting of an assembly of $N$
spins, each placed at a site of a $d$-dimensional Euclidean
lattice. Let lowercase italic letters, $i$, $j$,$\cdots$, denote
such sites, and $s_i$, $s_j$, $\cdots$, the relevant spin
variables, so that the energy pertaining to a given configuration
${\bf s}=\{s_i\}_{i=1}^N$ has the form
\begin{equation}
\label{E:nrg} E({\bf s})= -\frac{1}{2} \sum_{i=1}^N  s_i\,
\sum_{j\sim i} s_j,
\end{equation}
where the symbol $\sim$ in the second sum restricts it to the
sites $j$ adjacent to $i$.

As it is well known, the  Ising model\cite{A:Ising} is amenable of
an exact solution on simple two dimensional lattices
\cite{A:Onsa,B:2DIM}, whereas for more complex lattices or higher
dimensions the study of its thermodynamic properties strongly
relies on numeric simulations. According to the so-called {\it
dynamic Monte Carlo method}, the thermodynamic canonical average
of a generic observable $X({\bf s})$ is obtained as a simple
algebraic average over the Markov chain of configurations produced
by a suitable stochastic algorithm \cite{B:MC1}. Indeed each
configuration is obtained from the previous one so that, in the
asymptotic limit, the probability $P_T({\bf s})$ that a given
configuration $\bf s$ occurs at temperature $T$ is proportional to
the Boltzmann canonical factor, $\exp\left[-T^{-1}E({\bf
s})\right]$, independent of the initial configuration.

One of the most commonly exploited algorithms is  the Glauber's
single-spin-flip algorithm \cite{A:Glauber}. According to
Glauber's prescription two subsequent configurations can differ at
most by the value of a single spin variable. More precisely the
spin of the configuration $\bf s$ relevant to the site $k$ can
flip with probability
\begin{equation}
\label{E:Gp} p^{\rm G}_T({\bf s},k) = \frac{1}{1+\exp\left[T^{-1}
\Delta E_k({\bf s}), \right]}
\end{equation}
where
\begin{equation}
\Delta E_k({\bf s}) = 2 \, s_k \sum_{i \sim k}  s_i
\end{equation}
is the energy variation consequent to the process. Glauber's
dynamics is completely defined by Eq. (\ref{E:Gp}) together with a
prescription for updating the spin system. The latter is commonly
chosen in view of a computational optimization and typically
consists in an unphysical sweep along parallel lattice
lines\cite{N:US}.

As we mentioned above such a dynamics has a computational rather
than physical origin, having been devised as an efficient way to
give estimates of the canonical averages. Nevertheless it is
usually given a physical interpretation \cite{B:MC1}: the energy
variation consequent to the spin-flips realizing the evolution of
the system occur due to the coupling of the spin degrees of
freedom to a heat-bath at temperature $T$.

In the last decade a number of new dynamics for the Ising model
were introduced\cite{A:Mariz,A:Wang,A:Heri}, where the energy
variations are typically required to occur uniformly throughout
the whole sample.

Here we introduce an alternative dynamics exhibiting the diffusive
behaviour of a random walk in an evolving landscape generated by
the energy of the spin couplings in the real space. The spin-flips
are induced by a random walker hopping across the sites of the
underlying discrete structure. The motion of such a random walker
is affected by the spin interaction, being biased towards those
sites where a spin-flip is energetically more favorable. It should
be remarked that a relaxation dynamics occurring under the action
of random walkers was introduced by Creutz\cite{A:Creutz}, and
subsequently exploited by many authors. However our walkers act
quite differently from Creutz's {\it demons}. Indeed the latter
diffuse freely, unbiased by the magnetic configuration of the
system, and, in order to simulate the micro-canonical ensemble,
they induce a spin flip only if they can afford it, according to
the individual energy they are endowed with.

In fact our model in inspired by the physical behaviour of manganites
\cite{dagotto}, such as ${\rm La}_x {\rm Ca}_{1-x}{\rm Mn}{\rm O}_3$, where the 
spin dynamics is determined by the presence of diffusing excitations interacting 
with the magnetic degrees of freedom. In particular, for the manganites the 
excitations are given by the charged carriers, electrons and vacancies, present 
in the non-stoichiometrical 
compounds. Obviously, this model is not intended to be a phenomenological
description of such complex systems, but it aims to evidence the influence
of diffusing excitation interacting with spins on the thermodynamic behaviour of
the paradigmatic model of all magnetic phase transitions.

The analysis of the diffusive dynamics is carried out from the numerical point 
of view. All the simulations illustrated in the present paper refer to
two-dimensional arrays of spins, so that the observed features of
the statistical model can be compared to the analytically known
results pertaining to the canonical Ising model.

The plan of the paper is as follows: in Section \ref{S:DD} we
describe and briefly discuss the diffusive algorithm. Section
\ref{S:NCES} and \ref{S:CB} are devoted to the analysis of the
limit situation where the dynamics is realized by a single walker.
In the former we provide numerical evidences that the magnetic
system is driven to a thermodynamically well-behaved steady state
which differs from the canonical equilibrium state in a non
trivial-manner. Since a magnetic phase transition  is still
observed, the consequent critical behaviour is analyzed in Section
\ref{S:CB}. The estimated critical exponents do not deviate
significantly from the values pertaining to the canonical Ising
model. In Section \ref{S:DE} the results of simulations where the
system evolves subject to more than a single walker are analyzed.
Section \ref{S:CP} consists in our conclusions.

\section{Diffusive  Thermal Dynamics}
\label{S:DD} As we mentioned in the previous section the
relaxation dynamics we propose is realized by means of  random
walkers diffusing through the sites of the Ising system. The
probability that a walker  located at site $i$ hops on site $j$
and flips the relevant spin, where $j$ is one of the $2 d$
neighbours of $i$, is given by
\begin{equation}
\label{E:pij} p_T({\bf s},i,j)= (2 d)^{-1}  \, p_T^{\rm G}({\bf
s},j)
\end{equation}
Hence, for any configuration ${\bf s}$ of the Ising system and for
any value of the temperature $T$, the diffusion of the walker is
biased towards those sites where a spin-flip is more likely to
occur according to Glauber's probability, Eq. (\ref{E:Gp}). Note
that Equation (\ref{E:pij}) implicitly yields a probability
\begin{equation}
\label{E:pii} p_T({\bf s},i,i)= 1- (2 d)^{-1}\,\sum_{k \sim i}
p_T^{\rm G}({\bf s},k)
\end{equation}
that the walker does not move. If this is the case the magnetic
configuration of the system ${\bf s}$ remains unchanged as well.
According to equations (\ref{E:pij}) and (\ref{E:pii}) the
evolution of the probability $P({\bf s},i,t)$ that, at time step
$t$, the magnetic system is in the configuration ${\bf s}$ while
the walker is located at site $i$, is governed by the master
equation
\begin{eqnarray}
P_T({\bf s},i,t+1)-P_T({\bf s},i,t)=\nonumber\\
(2 d)^{-1}   \sum_{j \sim i}
\left[  P_T({\bf s}'_i,j,t)\, p_T^{\rm G}({\bf s}'_i,i) +  \right. \nonumber\\
\left.  - P_T({\bf s},i,t)\,p_T^{\rm G}({\bf s},j) \right],
\end{eqnarray}
where ${\bf s}'_k$ is a shorthand notation for the configuration
obtained flipping the spin $s_k$ of ${\bf s}$. Unfortunately the
master equation approach is of very little use in analyzing the
long-time behaviour of this process. Actually in the canonical
heat-bath case the master equation was exploited to build a
dynamics driving the system to a known asymptotic probability.
This was obtained by imposing the quite restrictive  {\it detailed
balance condition}. Conversely, we introduced an evolution
algorithm and our aim is the study of the resulting asymptotic
state of the Ising system, if any. Hence we carry out our analysis
mainly by means of numerical simulations.
 For a better comparison we focus on the two-dimensional system, where most of the results pertaining to the canonical Ising model are exactly known. Hence all of our simulations refer to an Ising system consisting of
$N=L^2$ spins placed at the sites of a two-dimensional square
lattice. In order to avoid time consuming procedures dealing with
the walkers bumping into the borders of the system we adopt
periodic boundary conditions.

In the next two Sections we focus on the situation where the
evolution of the Ising system is realized by a single walker. This
is a limit case, in that, for a reasonably macroscopic system, it
corresponds to a vanishing density of walkers. The results
pertaining to larger densities are discussed in Section
\ref{S:DE}.
\section{Non-Canonical Equilibrium States}
\label{S:NCES} 
\begin{figure}[h!]
\begin{center}
\includegraphics[width=8.6cm]{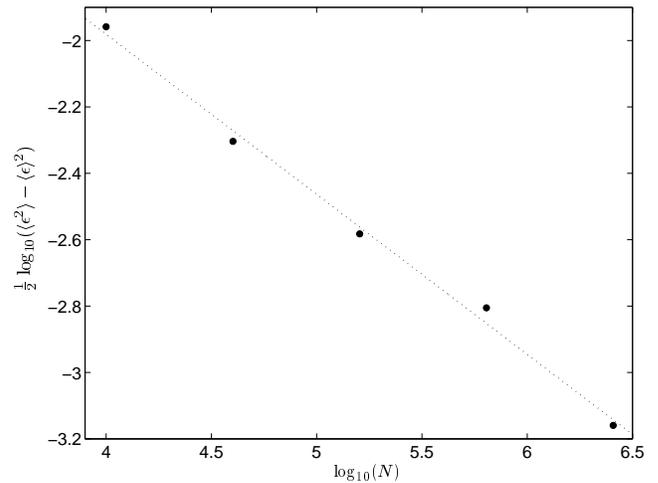}
\end{center}
\caption{Finite size scaling for the fluctuation about the average
value of the specific energy for an Ising system subject to the
diffusive dynamics described in Section \ref{S:DD} at $T=2.4$. The
slope of the linear fit (dotted line) of the measured data
($\bullet$),  $-0.48 \pm 0.02$, is in good agreement with the
expected value, $0.5$. All the measurements were carried out in
the stationary regime.} \label{F:nrgF}
\end{figure}

\begin{figure}[h!]
\begin{center}
\includegraphics[width=8.6cm]{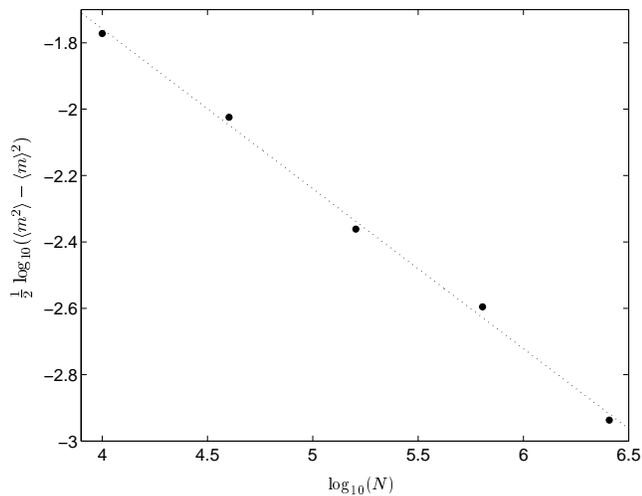}
\end{center}
\caption{Finite size scaling for the fluctuation about the average
value of the specific magnetization for an Ising system subject to
the diffusive dynamics described in Section \ref{S:DD} at $T=2.4$.
The slope of the linear fit (dotted line) of the measured data
($\bullet$),  $-0.48 \pm 0.02$, is in good agreement with the
expected value, $0.5$. All the measurements were carried out in
the stationary regime.} \label{F:magF}
\end{figure}

\begin{figure}[h!]
\begin{center}
\includegraphics[width=8.6cm]{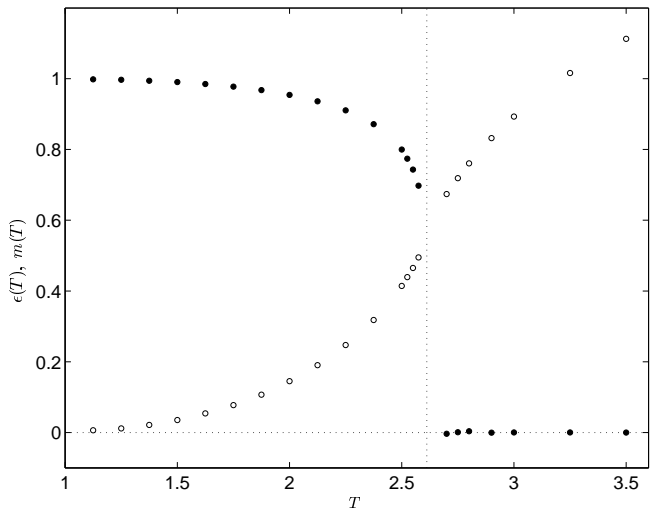}
\end{center}
\caption{Macroscopic observables for a 400 $\times$ 400 Ising
array. Filled circles ($\bullet$): specific magnetization; open
circles ($\circ$): specific energy.} \label{F:SpMag}
\end{figure}

\begin{figure}[h!]
\begin{center}
\begin{tabular}{cc}
\includegraphics[width=4cm]{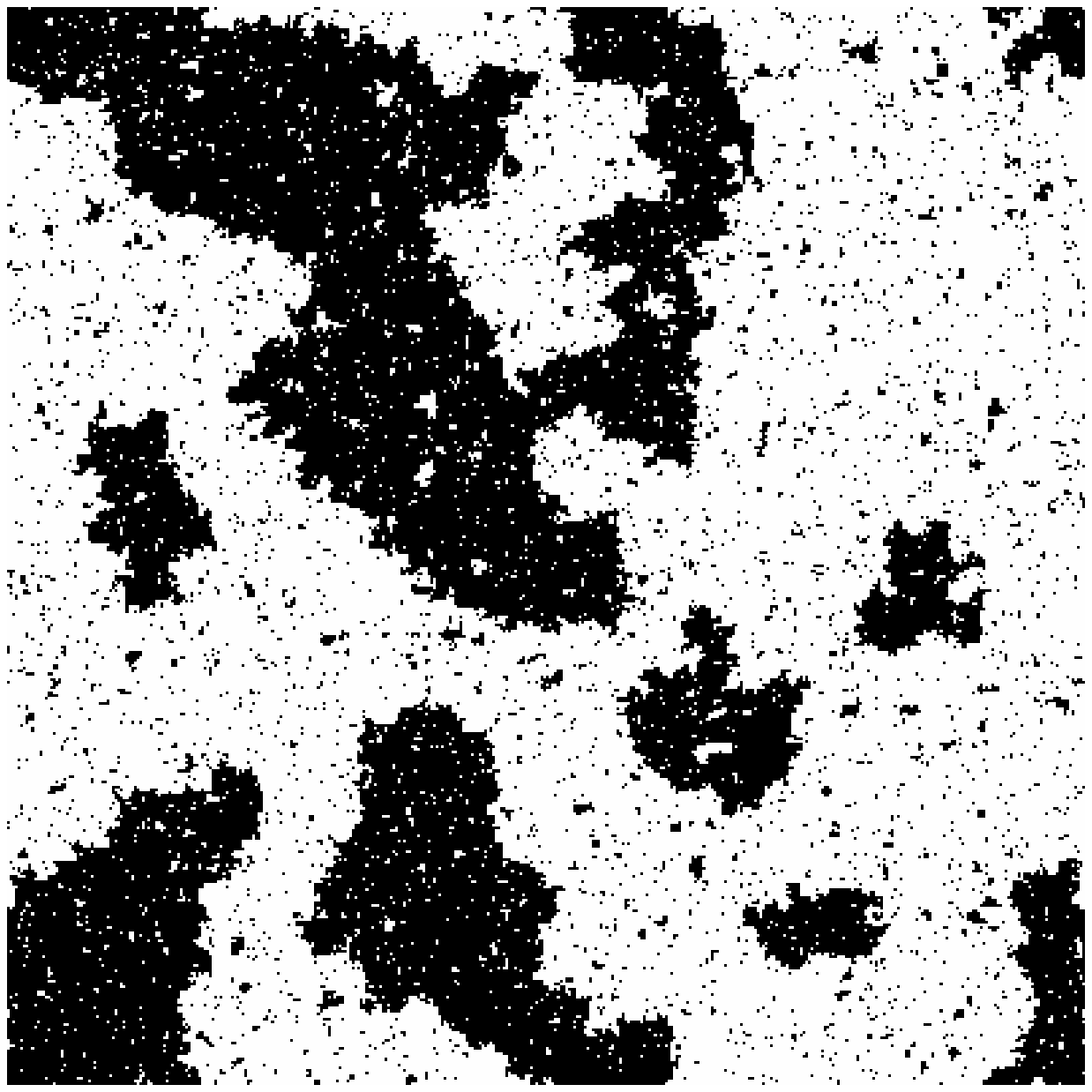} & \includegraphics[width=4cm]{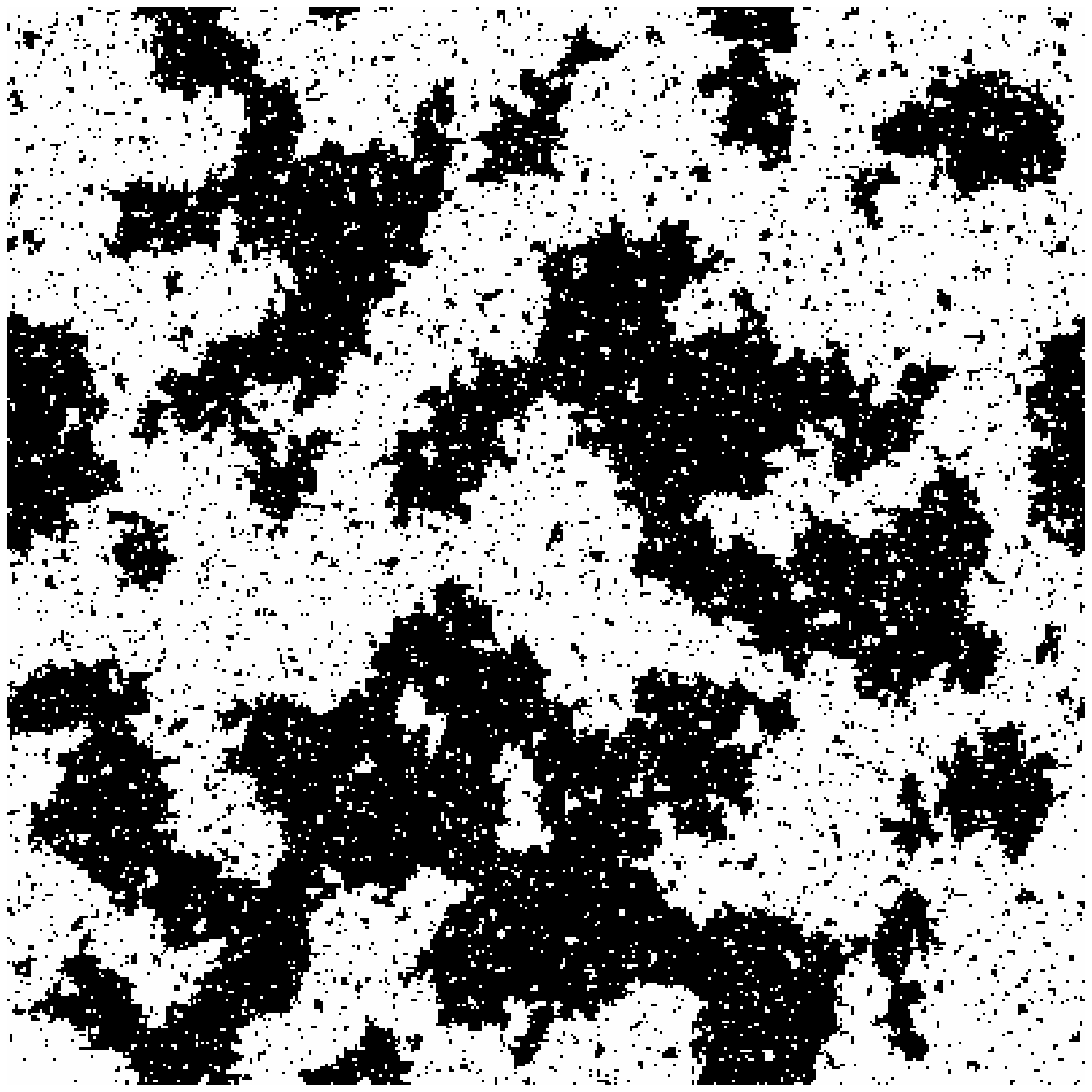}
\end{tabular}
\end{center}
\caption{Typical configurations for a 400 $\times$ 400 Ising array
with magnetization  $\langle m \rangle =0.86$  for the diffusive
dynamics (left panel) and the Glauber's dynamics (right panel).
Note that the same  magnetization  is attained for different
temperatures. More precisely  $T^{\rm diff}=2.4 > T^{\rm Ising}$
and  $T^{\rm Glauber}=2.13 < T^{\rm Ising}$, where $T^{\rm
Ising}_{\rm c}= 2/ {\ln} (1+\sqrt 2)\approx 2.269$ is the critical
temperature of the canonical Ising model. Note further that in the
diffusive case the domains present smoother boundaries. }
\label{F:domains}
\end{figure}

In the following we give numerical evidence that
the spin-flip dynamics realized by a single walker  drives the
magnetic system to a thermodynamically well-behaved steady state
which differs from the canonical equilibrium state in a non
trivial manner.
\begin{figure}[h!]
\begin{center}
\includegraphics[width=8.6cm]{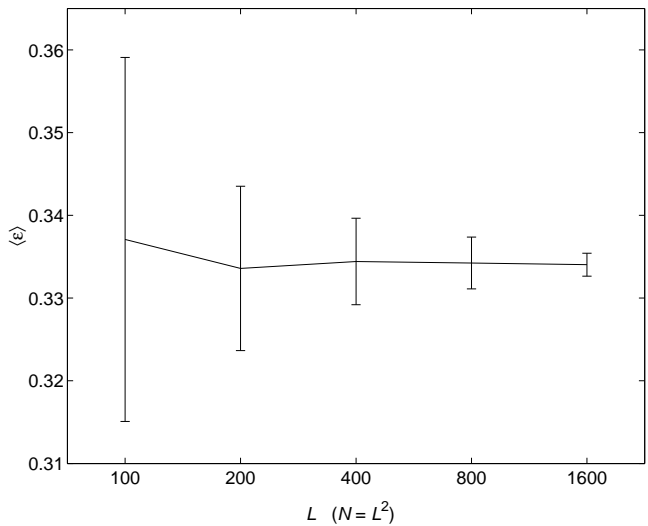}
\end{center}
\caption{Finite size scaling for the specific energy of an Ising
system subject to the diffusive dynamics described in Section
\ref{S:DD} at $T=2.4$. All the measurements were carried out in
the stationary regime. The error-bars represent the fluctuations
about the average values. } \label{F:nrgS}
\end{figure}
\begin{figure}[h!]
\begin{center}
\includegraphics[width=8.6cm]{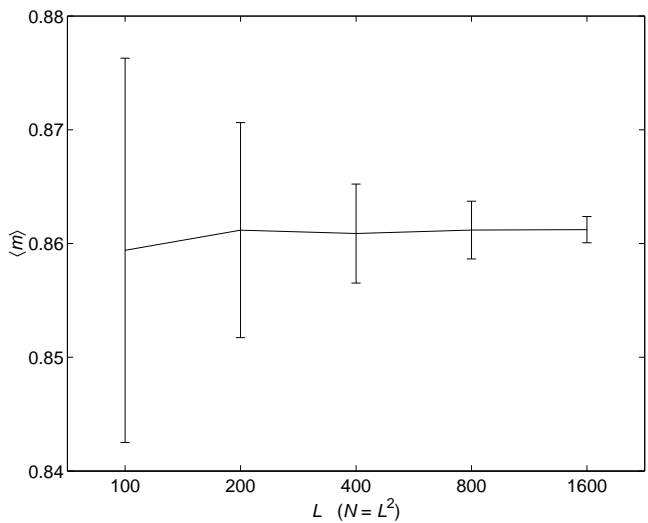}
\end{center}
\caption{Finite size scaling for the specific magnetization of an
Ising system subject to the diffusive dynamics described in
Section \ref{S:DD} at $T=2.4$. All the measurements were carried
out in the stationary regime. The error-bars represent the
fluctuations about the average values. Note that despite the
symmetry of the Hamiltonian the measured value for the specific
magnetization is non-zero. Similar to the canonical case the Ising
system displays a phase transition.} \label{F:magS}
\end{figure}
All the simulations we performed show that, at a given value of
the external parameter $T$ the system eventually reaches a steady
state, characterized by a well defined value of the (time) average
of the macroscopic observables. These features are clearly
recognizable in Figures \ref{F:nrgS} and \ref{F:magS}, where the
average values of the specific energy $\epsilon({\bf s}) = N^{-1}
E({\bf s})$ and magnetization $m({\bf s})= N^{-1} \sum_{i=1}^N
s_i$ are plotted for systems with different sizes, at a fixed
value of the temperature parameter. Furthermore, as one would
require, the entire sample and any reasonably macroscopic portion
of it are characterized by the same specific values of the
thermodynamic observables. As it is clearly shown in Figures
\ref{F:nrgF} and \ref{F:magF}, the fluctuations about the average
value of the specific thermodynamic observables exhibit the
expected scaling behaviour, decreasing as the inverse square root
of the system size.
Figure \ref{F:SpMag} shows the average values of the specific
energy and magnetization at different values of the temperature
parameter, $T=\beta^{-1}$. Note that, similar to the canonical
Ising model, the system exhibits a magnetic phase transition.
We verified that the long time behaviour of the macroscopic
observables of the system is uniquely determined by the only
external parameter characterizing the dynamics, namely the
temperature $T$. In particular different choices for the initial
configuration of the magnetic system do not affect the asymptotic
behaviour of the system, except for a possible trivial bias on the
orientation of the spontaneous magnetization. Similar to what
happens in the heat-bath case, both of the possible orientation of
the specific magnetization are in principle equally likely, but a
strongly magnetized initial configuration is very likely to evolve
in a stationary state exhibiting a spontaneous magnetization along
the same direction. When the initial configuration has no net
magnetization there is no bias on the direction of the spontaneous
magnetization exhibited by the system below the critical
temperature. In this situations the formation of macroscopic
domains exhibiting opposite net magnetization is observed (see
Fig.\ref{F:domains}). In the long time regime one of the domains
eventually prevails against the others.
As we already mentioned the system exhibits a magnetic phase
transition. From Fig. \ref{F:SpMag} and  \ref{F:domains}, it is
clear that the critical temperature for the diffusive dynamics is
significantly larger than the value pertaining to the canonical
Ising model. Such value,
 $T^{\rm Ising}_{\rm c}= 2/ {\ln} (1+\sqrt 2)\approx 2.269$
is analytically known and the heat-bath dynamics yield quite
satisfactory numerical estimates of it. As we will see in the next
Section, accurate estimates yield the value $T_{\rm c}= 2.612 \pm
0.002$ for the critical temperature of an Ising system evolving
under the action  of a single walker. However, such a quantitative
difference cannot be accounted for by means of  a simple rescaling
of the temperature. Indeed if the diffusive dynamics acted as
heat-bath dynamics with a rescaled temperature, $T'=\tau(T)$,  the
generic configuration $\bf s$ would occur with a probability
$P({\bf s},T) \propto \exp \left[-{T'}^{-1} E({\bf s})\right]$ in
the asymptotic regime \cite{B:MC1}. Hence the joint probability
for the occurrence of a configuration such that $\epsilon({\bf
s})=\epsilon'$ and $m({\bf s})=m'$ would be of the form
\begin{equation}
\label{E:jointP} {\cal P}(\epsilon', m', T)= Z(T) J(\epsilon', m')
e^{-{T'}^{-1} N \epsilon'},
\end{equation}
where
\begin{equation}
\label{E:combF}
 J(\epsilon', m') = \sum_{\bf s} \delta(\epsilon'-\epsilon({\bf s})) \,
\delta(m'-m({\bf s}))
\end{equation}
and
\begin{equation}
\label{E:partF} Z(T)= \sum_{m' \epsilon'} J(\epsilon', m')
\,e^{-{T'}^{-1}N \epsilon'} =
 \sum_{\bf s} e^{ -{T'}^{-1} E(\bf s)}
\end{equation}
Last equality of Eq. (\ref{E:partF}) was obtained making use of
Eq. (\ref{E:combF}). Now note that, according to Eqs.
(\ref{E:jointP}) and (\ref{E:combF}),
\begin{equation}
\tilde {\cal P}(\epsilon, m, T) \equiv \frac{{\cal P}(\epsilon, m,
T)}{\sum_m {\cal P}(\epsilon, m, T)} =\frac{J(\epsilon, m)}{\sum_m
J(\epsilon, m)}
\end{equation}

\begin{figure}[h!]
\begin{center}
\includegraphics[width=8.6cm]{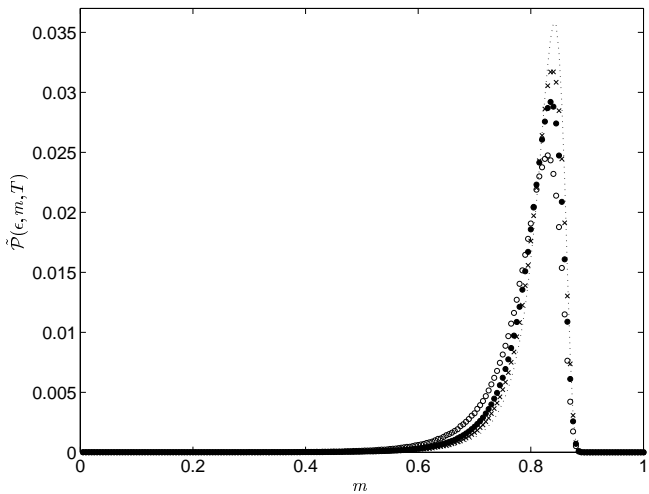}
\end{center}
\caption{Normalized joint distributions $\tilde {\cal
P}(\epsilon,m,T)$ for a $20 \times 20$ Ising array subject to the
diffusive dynamics. All of the curves refer to the specific energy
$\epsilon = 0.220$. Filled circles ($\bullet$): $T=2.25$; open
circles ($\circ$): $T=2.50$; crosses ($\times$): $T=2.75$; dotted
line: heat-bath dynamics (temperature independent).}
\label{F:dos1}
\end{figure}

\begin{figure}[h!]
\begin{center}
\includegraphics[width=8.6cm]{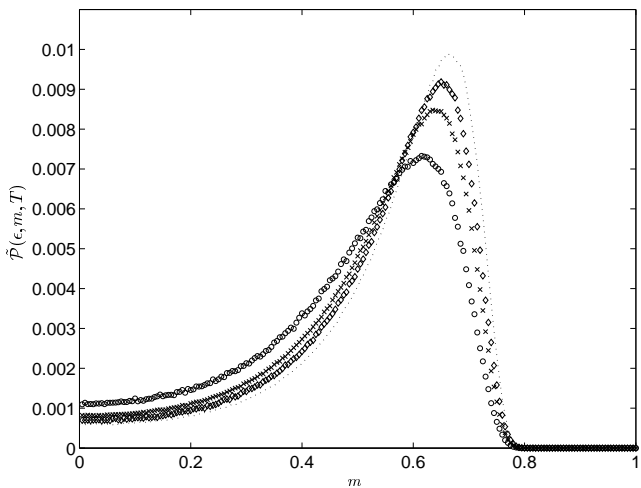}
\end{center}
\caption{Normalized joint distributions $\tilde {\cal
P}(\epsilon,m,T)$ for a $20 \times 20$ Ising array subject to the
diffusive dynamics. All of the curves refer to the specific energy
$\epsilon = 0.345$. Filled circles ($\bullet$): $T=2.50$; open
circles ($\circ$): $T=2.75$; crosses ($\times$): $T=3.00$; dotted
line: heat-bath dynamics (temperature independent).}
\label{F:dos2}
\end{figure}

Hence, if the rescaling hypothesis was true, the plot of $\tilde
{\cal P}(\epsilon, m, T)$ versus $m$ at a fixed value of the
specific energy would not depend on the temperature: the curves
pertaining to the same specific energy at different values of the
temperature $T$ would overlap. This can be verified with great
precision in the case of the heat-bath dynamics \cite{N:resc} The
numeric estimates of these curves for two different values of the
specific energy $\epsilon$ are plotted in figures \ref{F:dos1} and
\ref{F:dos2}.
The fact that the curves pertaining to the diffusive dynamics at
different temperatures are well distinct from one another and from
the (temperature independent) curve characterizing the canonical
Ising model proves that the differences between the diffusive and
the heat-bath dynamics cannot be accounted for by means of a
simple rescaling of the temperature.

In summary the steady state asymptotically reached by an Ising
system subject to the action of the diffusive dynamics is
thermodynamically well-behaved, and yet it is non-trivially
different from the canonical equilibrium of the Ising model. Hence
we will refer to the steady state of the diffusive dynamics as a
{\it non-canonical equilibrium state}.

In the following section we analyze the critical behaviour of the
Ising system subject to the diffusive dynamics.
\section{Critical Behaviour}
\label{S:CB} As we already observed in the previous Section, an
Ising system subject to a single  diffusive walker exhibits a
magnetic phase transition  at a finite value of the temperature
(see Figure \ref{F:SpMag}). Similar to what happens with the Ising
model, this phenomenon is accompanied by a singular behaviour of
the thermodynamic functions.
\begin{figure}[!ht]
\begin{center}
\includegraphics[width=8.6cm]{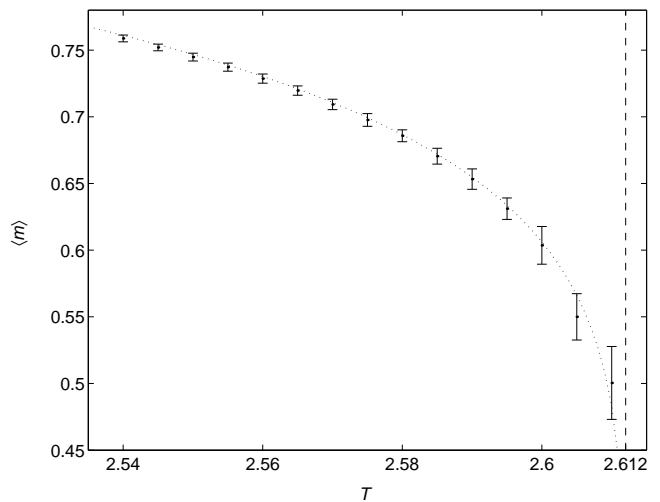}
\caption{Critical exponent, $\beta$, of the magnetization for an
Ising system subject to the diffusive dynamics. The measures were
performed on a system consisting of $N=2000\times2000$ spins. The
error bars denote the standard deviation of the measures. The
dotted line is the best fit, $y=A\,|T-T_{\rm c}|^\beta$. The
estimated values for the critical temperature and for the exponent
are  $T_{\rm c}=2.612 \pm 0.001$ and $\beta=0.127\pm 0.002$,
respectively. The latter is consistent with the relevant critical
exponent of the canonical Ising model, $\beta_{\rm Ising}=1/8$.
The vertical dashed line indicates the critical temperature.}
\label{F:beta}
\end{center}
\end{figure}
In Figure \ref{F:beta} accurate data for the magnetization of an
Ising system subject to the diffusive dynamics are plotted. The
error bars represent the standard deviations about the average
values. Similar to the canonical case, the data show a good
agreement with a critical behaviour of the form $m(T)\sim
|T-T_{\rm c}|^\beta$, where the estimated values are for the
critical temperature and for the critical exponent are $T_{\rm
c}=2.612\pm0.001$ and $\beta=0.127\pm0.02$, respectively. The
latter result is in good agreement with the relevant critical
exponent of the two-dimensional canonical Ising model, $\beta_{\rm
Ising}=1/8$. As we already discussed in the previous Section, the
critical temperature is appreciably larger than the canonical
value, $T_{\rm c}^{\rm Ising}\approx 2.269$.

\begin{figure}[!ht]
\begin{center}
\includegraphics[width=8.6cm]{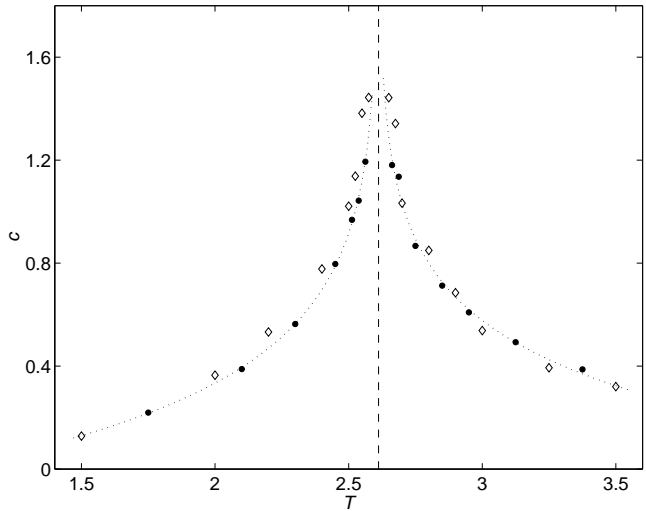}
\caption{Specific heat for an Ising system subject to diffusive
dynamics ($\bullet$). The dotted curves fitting the data are of
the form $f(T)=a \, +b\,\log(|T-T_{\rm c}|)$. The vertical dashed
line indicates the estimated value of the critical temperature,
$T_{\rm c}=2.612\pm 0.002$. The diamonds ($\diamond$) refer to the
quantity $ c_{\rm f}(T)$ defined in
 Eq. (\ref{E:flC}). }
\label{F:SpH}
\end{center}
\end{figure}

\begin{figure}[h!]
\begin{center}
\includegraphics[width=8.6cm]{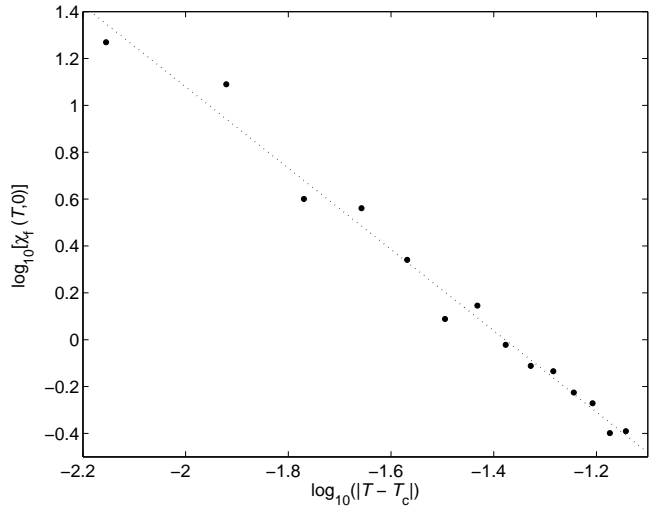}
\caption{Log-log scale plot of the magnetic susceptibility
$\chi_{\rm f}$ (defined in terms of the magnetic fluctuations, see
Eq. (\ref{E:chif})) versus $|T-T_{\rm c}|$. The estimated slope of
the straight line fitting the data is $-1.73\pm0.07$. This value
is consistent with the critical behaviour of the magnetic
susceptibility for the two-dimensional canonical Ising model,
characterized by the critical exponent $\gamma_{\rm Ising}=7/4$.}
\label{F:gamma}
\end{center}
\end{figure}

Figure \ref{F:SpH} shows the specific heat $c(T)=d\,
\langle\epsilon\rangle_T /d\,T$ for a $400 \times 400$ Ising array
subject to the diffusive dynamics as a function of the temperature
$T$. The vertical dashed lines are placed at the critical value of
the
 temperature. The dotted curves refer to functions of the form
$f(T)=a \, +b\,\log(|T-T_{\rm c}|)$, and they fit quite
satisfactorily the
 data. Hence, similar to the canonical case, there is a strong
signature of a logarithmic divergence of the specific heat at the
critical temperature.
Figure \ref{F:gamma} shows a log-log scale plot of the quantity
\begin{equation}
\label{E:chif}
 \chi_{\rm f}(T)\equiv
\frac{N}{T} \left[\langle m^2\rangle_T- \langle
m\rangle_T^2\right]
\end{equation}
versus $|T-T_{\rm c}|$. The data are consistent with a power law
of the form $\chi_{\rm f}(T)\sim |T-T_{\rm c}|^\gamma$, and the
slope $\gamma=1.73\pm0.06$ of the linear fit is in good agreement
with the critical exponent governing the  behaviour of the same
quantity in the of the  canonical case,  $\gamma_{\rm Ising}=7/4$.
We recall that for a system at canonical equilibrium the
fluctuation-dissipation theorem yields the relations $c(T)= c_{\rm
f}(T)$ and $\chi(T)= \chi_{\rm f}(T)$, where
\begin{equation}
\label{E:flC} c_{\rm f}(T) = \frac{N}{T^2}
\left[\langle\epsilon^2\rangle_T-
\langle\epsilon\rangle_T^2\right].
\end{equation}
and $\chi(T)=\partial\,\langle m\rangle_{T,h}/\partial\,h$ is the
magnetic susceptibility of the system. In our case this is not
necessarily true, since the equilibrium distribution for a
diffusive dynamics at the temperature $T$ is not proportional to
the canonical Boltzmann factor $\exp\left[-T^{-1} H\right]$, as we
discussed in the previous section. Hence the critical behaviour of
two quantities canonically related by a fluctuation-dissipation
relation should be explicitly analyzed and compared. A preliminary
step in this sense is shown in figure \ref{F:SpH}, where the
diamonds represent the estimates of the quantity $ c_{\rm f}(T)$
defined in Eq. (\ref{E:flC}). These results seem to indicate that
the diffusive character  of the dynamics does not produce large
deviations from the fluctuation-dissipation relation. Of course
this result needs to be checked through the analysis of the
response of the system to an external magnetic field. However we
mention that other studies \cite{A:NLSS1} of Ising systems subject
to dynamics which do not yield the canonical equilibrium state
actually estimate the critical exponent $\gamma$ in terms of the
quantity $ \chi_{\rm f}(T)$ defined in Eq. (\ref{E:chif}).

\begin{figure}[h!]
\begin{center}
\includegraphics[width=8.6cm]{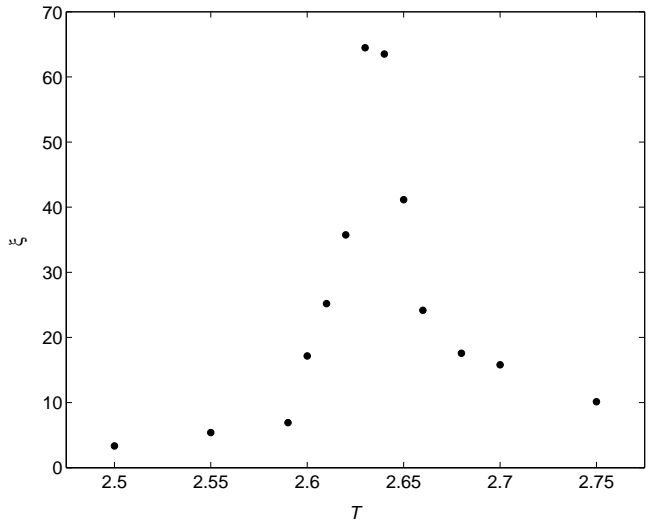}
\caption{Correlation length as a function of the temperature for a
$600 \times 600$ Ising array subject to the diffusive dynamics.
Due to finite size effects, the critical behaviour is exhibited
for temperature slightly higher than the estimated critical value,
$T_{\rm c}=2.612 \pm 0.002$.} \label{F:xiT}
\end{center}
\end{figure}

For temperatures sufficiently far from the critical value the
correlation function
\begin{equation}
\label{E:Gamma} \Gamma_T(i,j) = \langle s_i \, s_j\rangle_T -
\langle s_i \rangle_T \, \langle s_j\rangle_T
\end{equation}
exhibits an exponential decrease as the distance between the
relevant sites , $r_{i\,j}=||i-j||$, increases:
\begin{equation}
\Gamma_T(i,j) \propto \exp \left[ - \frac{r_{i\,j}}{\xi(T)}\right]
\end{equation}
As it is shown in Fig. \ref{F:xiT} the correlation distance
$\xi(T)$ features a maximum very close to the critical
temperature. The fact that this maximum suggests a critical
temperature slightly larger than the value previously (and more
accurately) estimated must be ascribed to finite size effects. We
further mention that the correlation lengths are roughly
consistent with a power-law divergence of the form $\xi(T) \sim
|T~-~T_{\rm c}|^{-\nu}$, the estimated exponent being remarkably
close to the relevant critical exponent of the canonical
two-dimensional Ising model, $\nu_{\rm Ising}=1$.
As it is well known \cite{A:Kada}, for temperatures very close to
the critical value, and hence for correlation lengths very large
compared to the lattice constant, the correlation function of the
canonical Ising model is characterized by a power-law behaviour of
the form $\Gamma_T(i,j) \propto r_{i\,j}^p$. The relevant critical
exponent is defined to be $\eta= p-d+2$, where $d$ is the
dimensionality of the Ising array. For the two-dimensional
canonical Ising model its value is exactly known to be $\eta_{\rm
Ising}=1/4$. Such behaviour is expected to be observable in the
distance range $1\ll r_{\i\,j} \ll \xi(T)$.
\begin{figure}[h!]
\begin{center}
\includegraphics[width=8.6cm]{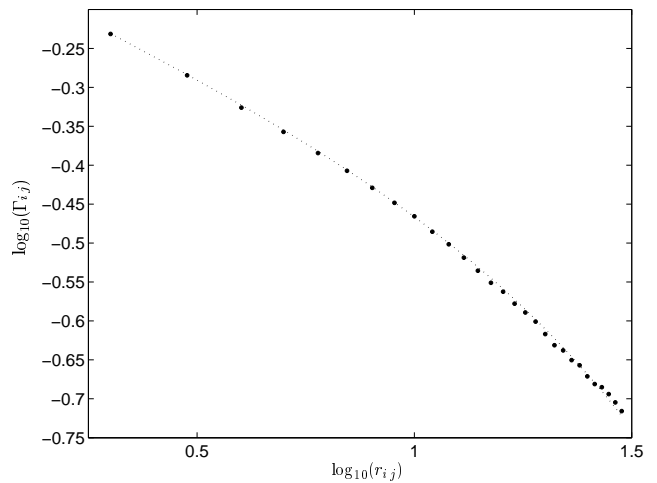}
\caption{Log-log scale plot of the correlation function
$\Gamma_{T=2.63}(i,j)$ versus the inter-spin distance, $r_{i\,j}$.
The data appreciably deviate from a power law behaviour. The
dotted line refer to a function of the form defined by Eq.
(\ref{E:CFfit}). The exponent $\eta$ is quite close to the
critical exponent $\eta_{\rm Ising}=1/4$. More precisely we
estimate $\eta_{2.63}=0.252\pm0.002$. The same fit satisfactorily
applies to the data relevant to $T=2.64$, where the exponent
$\eta_{2.64}=0.266\pm 0.003$ is once again in good agreement with
the canonical value.} \label{F:eta}
\end{center}
\end{figure}
As it is shown in Figure \ref{F:eta} the data yielding the largest
correlation length are not consistent with a pure power law
behaviour, not even for relatively small distances. This suggests
that the relevant
 temperatures, $T=2.63$ is not sufficiently close to the critical value.
 Moreover it should be recalled that the finite size of
the system can introduce appreciable deviations from the
theoretical prediction, which is strictly true in the
thermodynamic limit. Nevertheless we mention that a function of
the form
\begin{equation}
\label{E:CFfit} f(r) = a\, \exp(-\frac{r}{\xi(T)})\, r^{- \eta},
\end{equation}
fits the data quite satisfactorily over a wide range of distances.
The estimate for the exponent, $\eta = 0.252 \pm 0.002$ is
remarkably close to the critical exponent $\eta_{\rm Ising}$ of
the two-dimensional Ising model. A fit of the same kind
satisfactorily applies to the data relevant to the temperature
$T=2.64$, and it gives a critical exponent $\eta = 0.266 \pm
0.002$ once again in good agreement with the canonical value.
 These results could be an indication of the fact that, in the suitable
regime, the correlation function $\Gamma_{T\approx T_{\rm
c}}(i\,j)$ actually exhibits a power law behaviour characterized
by the same critical exponent as the two-dimensional canonical
Ising model.

The results illustrated so far suggest that a two-dimensional
Ising ferromagnet subject to the diffusive dynamics exhibits a
critical behaviour belonging to the same universality class as the
canonical case.

 Before concluding the present section we mention a further result which
has no counterpart in the canonical case. Indeed, in the diffusive
case, it is possible to relate the configuration of the magnetic
system to the position of the random walker causing its evolution.
More precisely it is possible to measure to what extent the
presence of the random walker at a given site $i$ and the local
magnetization at a given site $j$ influence each other. Numerical
estimates of such correlation, which we illustrate elsewhere
\cite{A:forthcoming}, show that it exhibits a critical behaviour
in correspondence of the critical temperature.

\section{Recovering the Canonical Equilibrium: Density of Walkers}
\label{S:DE} The results illustrated in the previous two Sections
refer to the case of an Ising system subject to the action of a
single walker. Here we analyze the results of simulations where
the evolution of a square Ising array consisting of $N$ spins is
realized by $n$ non interacting walkers. The simulation is
initialized giving each walker a randomly chosen position. Two or
more walkers are allowed to occupy the same site of the system.
Actually this is very likely to happen for a sufficiently large
density of walkers $\rho=n/N$ . An elementary step of the
simulation consists in the application of the algorithm described
in Section \ref{S:DD} to all of the walkers, according to a fixed
progressive sequence. When its turn comes, a given walker acts as
if the other ones were not present. A qualitative argument leads
us to expect that in the case of an infinite density of walkers
the results pertaining to the canonical Ising model are recovered.
Consider a very large density of walkers, such that $n \gg N$. In
this situation the effect of a single step of the simulation is
very similar to the action of the Glauber's dynamics with a random
sequence update. Indeed the probability that a walker flips a
given spin $s_i$ is given by the product of the probability $4/N$
that it is located at one of the four neighbours of site $i$ times
the hopping probability, $p^{\rm G}({\bf s}, i)/4$. But $p^{\rm
G}({\bf s}, i)/N$  is exactly the probability that the spin $s_i$
is flipped at any step of the Glauber's heat-bath dynamics with
random site update. Hence, in the limit case of an infinite
density of walkers, the diffusive dynamics should necessarily
drive the system to  the same asymptotic state as the heat-bath
dynamics, thus reproducing the results of the canonical Ising
model. The same should be clearly true for a sufficiently large
density of walkers, provided  that the consequent number of
elementary steps of the heat-bath dynamics drive the system
sufficiently close to the canonical equilibrium state.
\begin{figure}[h!]
\begin{center}
\includegraphics[width=8.6cm]{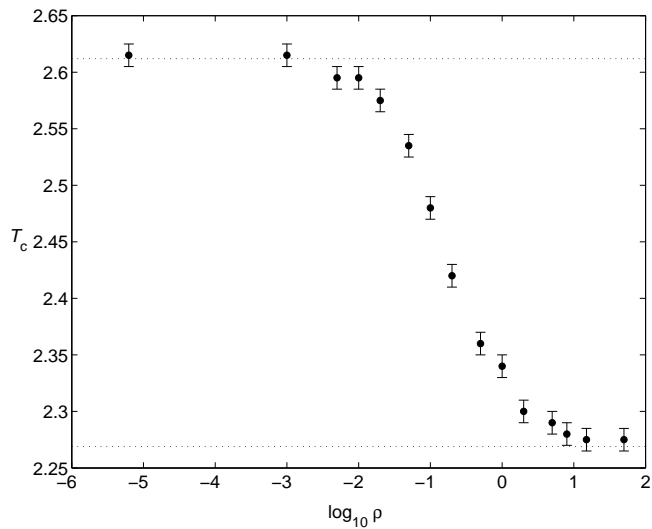}
\caption{Critical temperatures with increasing density of walkers
for a $400 \times 400$ Ising array. The horizontal dotted lines
are placed at the critical temperatures of the diffusive dynamics
with a single walker ($T_{\rm c}\approx 2.612$) and at the
critical temperature of the canonical Ising model ($T^{\rm
Ising}_{\rm c}\approx 2.269$).} \label{F:nTc}
\end{center}
\end{figure}
The numerical simulations we performed evidenced that the
diffusive dynamics drives the Ising system to a thermodynamically
well behaved stationary state, so that the average values of the
specific energy and magnetization are characterized only by the
temperature, $T$, and the density of walkers, $\rho$. In
particular they confirmed the qualitative argument discussed
above: the results pertaining to the canonical Ising model at a
given temperature $T$ were recovered for a sufficiently large
density of walkers. More generally we verified that the average
value of a macroscopic observable  relevant to a finite density
lies within the interval between the values pertaining to the
limit situations, namely the vanishing density limit $\rho =
N^{-1} \sim 0$ analyzed in Sections \ref{S:NCES} and \ref{S:CB},
and the canonical Ising model. The influence of the density of
walkers is synthesized in Figure \ref{F:nTc}, where the estimated
values of the critical temperature are plotted. Note that the
larger variation occurs for densities between $10^{-2}$ and  $1$.

\section{Conclusions and Perspectives}
\label{S:CP}
In this paper we exploited the basic microscopic
model of diffusion, namely the random walk, to build a dynamics
for the Ising ferromagnet which retains the diffusive character of
the thermal motion of a localized excitation. In this framework
the density of walkers is related both to the intensity and
spatial extent of the coupling of the spins to the thermal degrees
of freedom of the system, and hence to the velocity of the thermal
dynamics. The larger is the density of walkers, the faster is the
dynamics and the closer to the uniform heath-bath are its effects.

Even if our model is rather simplified with respect to the real physical systems
which inspired it, it exhibits new important qualitative features which should 
be relevant even in more complex and realistic situations. Therefore
it would be of great interest to verify whether the diffusive
thermal dynamics gives a satisfactory description for the slow
evolution of a magnetic system with hopping excitations. This
would of course require the extension of the study of its effect
on the more realistic three dimensional lattice. In view of
recognizing the action of a diffusive dynamics in a real system,
the results of Section \ref{S:CB} suggest that the study of the
critical behaviour is not particularly significant. Indeed the
latter does not allow to discriminate among the canonical and the
diffusive dynamics.

On the other hand the results illustrated in section \ref{S:DE}
indicate that the velocity of the non-equilibrium dynamics
(related to the degree of coupling of the spins to the thermal
vibrations, i.e. to the density of excitations induced in the
system by an external source) significantly affects the critical
temperature of the system. Hence the same system subject to
diffusive thermal dynamics with different velocity should display
different critical temperatures. Furthermore we remark that Fig.
\ref{F:domains} suggests that the velocity of the dynamics
influences the shape and growth of the macroscopic magnetic
domains.

 In particular a very slow diffusive dynamics gives rise to domains with smoother boundaries. This
feature suggests that the final shape of the magnetic domains
within a physical system possibly governed by a diffusive thermal
dynamics could be controlled by a fine tuning of the density of
excitations mediating the coupling to the thermal degrees of
freedom. A detailed study of domain growth in the diffusive
dynamics will be the subject of a forthcoming paper
\cite{A:forthcoming}.

\acknowledgements

We are grateful to R. De Renzi for useful discussions.

\bibliography{}

\begin{thebibliography}{abcdef 99}
\bibitem{N:resc} In that case the rescaling is trivial: $\tau(x)=x$.
\end{thebibliography}

\end{document}